\documentclass[preprint,noshowpacs,noshowkeys,aps,pra,nofootinbib]{revtex4-2}

\usepackage{graphicx}
\usepackage{amssymb}
\usepackage{amsmath}
\usepackage{stmaryrd}
\usepackage{float}
\usepackage{makecell}

\usepackage{pgf, tikz}
\usetikzlibrary{positioning}
\usepackage[simplified]{pgf-umlcd}

\newcommand{\Software}{\textsc{WaveTrain} }

\begin{document}

\title{WaveTrain: A Python package for numerical quantum mechanics of chain-like systems based on tensor trains}

\author{Jerome Riedel}
\email{jerome.riedel@fu-berlin.de}
\affiliation{
Institut f\"{u}r Chemie, Freie Universit\"{a}t Berlin\\ Altensteinstraße 23A, D-14195 Berlin, Germany}

\author{Patrick Gel\ss{}}
\email{p.gelss@fu-berlin.de}
\affiliation{
Institut f\"{u}r Mathematik, Freie Universit\"{a}t Berlin \\ Arnimallee 3--9, D-14195 Berlin, Germany}
\affiliation{Zuse-Institut Berlin, Takustraße 7, D-14195 Berlin, Germany}

\author{Rupert Klein}
\email{rupert.klein@math.fu-berlin.de}
\affiliation{
Institut f\"{u}r Mathematik, Freie Universit\"{a}t Berlin \\ Arnimallee 3--9, D-14195 Berlin, Germany}

\author{Burkhard Schmidt}
\email{burkhard.schmidt@fu-berlin.de}
\affiliation{
Institut f\"{u}r Mathematik, Freie Universit\"{a}t Berlin \\ Arnimallee 3--9, D-14195 Berlin, Germany}
\affiliation{Weierstra\ss -Institut f\"{u}r Angewandte Analysis und Stochastik, Mohrenstra\ss e 39, 10117 Berlin, Germany}

\date{\today}

\begin{abstract}
\Software is an open-source software for numerical simulations of chain-like quantum systems with nearest-neighbor (NN) interactions only.
The Python package is centered around tensor train (TT, or matrix product) format representations of Hamiltonian operators and (stationary or time-evolving) state vectors.
It builds on the Python tensor train toolbox \textsc{Scikit\_tt}, which provides efficient construction methods and storage schemes for the TT format.
Its solvers for eigenvalue problems and linear differential equations are used in 
\Software for the time-independent and time-dependent Schr\"odinger equations, respectively.
Employing efficient decompositions to construct low-rank representations, the tensor-train ranks of state vectors are often found to depend only marginally on the chain length $N$.
This results in the computational effort growing only slightly more than linearly with $N$, thus mitigating the curse of dimensionality.
As a complement to the classes for full quantum mechanics, \Software also contains classes for fully classical and mixed quantum-classical (Ehrenfest or mean field) dynamics of bipartite systems.
The graphical capabilities allow visualization of quantum dynamics ‘on the fly’, with a choice of several different representations based on reduced density matrices.
Even though developed for treating quasi one-dimensional excitonic energy transport in molecular solids or conjugated organic polymers, including coupling to phonons, \Software can be used for any kind of chain-like quantum systems, with or without periodic boundary conditions, and with NN interactions only.  

The present work describes version 1.0 of our \textsc{WaveTrain} software, based on version 1.2 of \textsc{scikit\_tt}, both of which are freely available from the GitHub platform where they will also be further developed.
Moreover, \Software is mirrored at SourceForge, within the framework of the \textsc{WavePacket} project for numerical quantum dynamics. 
Worked-out demonstration examples with complete input and output, including animated graphics, are available.
\end{abstract}

\maketitle

\section{Introduction}
\label{sec:intro}
Progress in ultra-fast experimental techniques, in particular the generation of ultra-short, intense laser pulses, has led to substantial advances in atomic and molecular physics, chemical reaction dynamics, material sciences and related fields~\citep{Schryver2001}. 
This has also motivated research in theoretical and simulation studies of quantum dynamics in recent years~\citep{May2000,Tannor2004,Grossmann2008}.
However, in marked contrast to electronic structure theory where a number of software packages have been under constant development for years or even decades and which have reached a remarkable degree of sophistication, general-purpose simulation software for quantum dynamics is relatively scarce. 
For example, QuTiP is an open-source Python framework for the dynamics of open quantum systems~\citep{Johansson2012,Johansson2013}.
Another framework for closed and open quantum systems, coded in Matlab, aims at applications in quantum optics and condensed matter~\citep{Norambuena2020}.
Furthermore, Libra offers a toolbox for quantum and classical dynamics simulations, including non-adiabatic processes in molecular system~\cite{Akimov2016}.
This is also a main feature of \textsc{WavePacket}, a general purpose package for solving coupled Schr\"{o}dinger or Liouville-von Neumann equations of closed and open quantum systems, respectively~\cite{Schmidt2017,Schmidt2018,Schmidt2019}.
Additionally, it offers modules for fully classical and mixed quantum-classical dynamics on an equal footing, as well as a module for optimal control.
The latter is the focus of QEngine~\citep{Sorensen2019}, a C++ library, and of Krotov, a Python implementation of quantum optimal control~\citep{Goerz2019}.

Quantum dynamical simulations using any of the software packages mentioned above are limited to rather few degrees of freedom.
This is because of the use of conventional grid techniques for the representations of quantum states and operators, thus suffering from the \textit{curse of dimensionality}, i.~e., the exponential growth of storage and CPU time with the number of dimensions.
One way to overcome this problem is the 
multi-configurational time-dependent Hartree (MCTDH) implementation and its multi-layer (ML) extensions~\citep{Beck2000,Meyer2009}.
This package is frequently used for complex quantum molecular dynamics simulation tasks, and it has evolved into a quasi-standard in the chemical physics community.
From the quantum physics point of view, similar concepts are formulated in terms of tensor networks.
In fact, it is well established that the (ML-)MCTDH algorithm corresponds to (hierarchical) Tucker tensor formats.
For various types of tensor networks, the ITensor software library is available for practical calculations~\citep{Fishman2020}.
In particular, it contains the density matrix renormalization group (DMRG) algorithm for computing low-energy states of quantum systems~\citep{Paeckel2019}.

The present work deals with high-dimensional quantum dynamics using tensor train (TT) representations of quantum states and operators, also known as matrix product states (MPS) and operators (MPO)~\cite{Affleck1987, Oseledets2009a, Oseledets2009b}. 
The idea behind this format is to decompose a high-dimensional tensor into a chain-like network of lower-dimensional tensors which enables us to simulate and analyze large-scale problems if the underlying coupling structure allows for low-rank decompositions. 
Several applications of tensor trains -- which can be considered as a special case of the ansatz used in the multi-layer (ML) variant of MCTDH~\citep{Beck2000,Meyer2009} mentioned above -- and tensor-train operators have shown that it is possible to mitigate
the curse of dimensionality and to tackle high-dimensional problems which cannot be solved using conventional numerical methods, see, e.g., dynamical systems~\cite{Klus2018, Luecke2021}, system identification~\cite{Gelss2019, Goessmann2020}, quantum mechanics~\cite{Veit2017, Gelss2022a, Gelss2022b}, and also quantum machine learning~\cite{Klus2019, Huggins2019}. 
Typically, the applications require the approximation of the solutions of systems of linear equations, eigenvalue problems, ordinary/partial differential equations.
For this reason, we use the open-source toolbox \textsc{Scikit-TT}\footnote{\url{https://github.com/PGelss/scikit_tt}}, a general-purpose package for tensor trains written in Python based on NumPy and Scipy.
It provides a powerful TT class as well as different modules for the automatic construction of tensor trains.
Furthermore, \textsc{Scikit-TT} comprises different solvers for algebraic problems which we need for our simulations.

Herein, we present version 1.0.0 of the \Software software package which specializes on high-dimensional quantum dynamics for systems with a chain-like topology and nearest-neighbor (NN) interactions only.
Using tensor-train (TT) representations based on the so-called SLIM scheme~\citep{Gelss2017}, this packet builds on \textsc{Scikit-TT} thus providing efficient low-rank tensor approximation approaches which aim at reducing the exponential scaling of the computational effort for solving time-independent and time-dependent Schr\"{o}dinger equations in many dimensions. 
Being restricted to the SLIM scheme for TT representations for chain-like quantum systems with NN interactions, this approach is less general than other tensor schemes such as the (hierarchical) Tucker format underlying the ML--MCTDH scheme, but has the advantage of very favorable scaling of the numerical effort with the chain length.

In our previous papers, the TT scheme was applied to the solution of the time-independent (TISE) and time-dependent Schr\"{o}dinger equation (TDSE) for exciton-phonon systems of NN type, i.e., quasi-1D excitonic chains, ranging from few to about one hundred sites~\citep{Gelss2022a,Gelss2022b}.
% far more than is feasible with most full quantum techniques 
There it was demonstrated that the storage consumption of the SLIM scheme scales linearly with the number of sites, and the scaling of the CPU time is only slightly less favorable. 
Moreover, for the case of the TISE, convergence with regard to the tensor rank was shown to be essentially independent of the system size. 
In another recent study, the efficiency in calculating ground states of chains of linear rotors interacting through their dipole moments was investigated. 
There, it was found that for these systems a TT-based approach is less time- and memory-consuming than the state-of-the-art implementation of ML-MCTDH~\cite{Mainali2021,Serwatka2022}.
Finally, it is mentioned that the \Software platform also contains modules for fully classical and hybrid quantum-classical dynamics dynamics, both for reference and/or for treating systems that are too complex for fully a quantum-mechanical treatment. 

\section{Physical systems and Hamiltonians}
\label{sec:ham}
\subsection{Tensor trains and the SLIM decomposition}
\label{sec:ham:slim}

Throughout the \Software software package we limit ourselves to the treatment of physical/chemical systems with a chain-like topology with NN (nearest neighbor) interactions only.
For such systems, quantum-mechanical Hamiltonians $H$ can be decomposed into operators that either act locally on single sites or that couple NN pairs in a chain with $N$ sites.
Using a so-called \emph{SLIM decomposition}~\cite{Gelss2017} where the origin of the acronym is due to the quantities $S,L,I,M$), the \emph{canonical representation} of the tensor $H \in \mathbb{R}^{(d_1 \times d_1) \times \dots \times (d_N \times d_N)}$ only consists of elementary tensors, where at most two (adjacent) components are unequal to the identity matrix: 
\begin{equation}
    \begin{split}
        H &= S_{1} \otimes I_2 \otimes \dots \otimes I_N \quad + \quad \dots \quad + \quad I_1 \otimes \dots \otimes I_{N-1} \otimes S_{N} \\
        & \quad + \quad \sum_{\lambda=1}^{\xi_1}  L_{1,\lambda} \otimes M_{2,\lambda} \otimes I_3 \otimes \dots \otimes I_N \quad + \quad \dots \\
				& \quad + \quad \sum_{\lambda=1}^{\xi_{N-1}}  I_1 \otimes \dots \otimes I_{N-2} \otimes L_{N-1,\lambda} \otimes M_{N,\lambda}\\
        & \quad + \quad \sum_{\lambda=1}^{\xi_N} M_{1,\lambda} \otimes I_2 \otimes \dots \otimes I_{N-1} \otimes L_{N,\lambda}. 
    \end{split}
		\label{eq:SLIM_1}
\end{equation}
Here all components $S_{i}$, $L_{i,\lambda}$, and $M_{i,\lambda}$ as well as the identities $I_i$ are matrices in $\mathbb{R}^{d_i \times d_i}$ where the $d_i$ are the dimensions of the Hilbert spaces characterizing quantum states on the sites $i$. 
Note that the last line of Eq.~(\ref{eq:SLIM_1}) is only to comply with periodic boundary conditions of cyclic systems and can be omitted otherwise. 

As shown in \cite{Gelss2017}, the structure of such a Hamiltonian corresponds to the topology of a tensor train (TT, also known as matrix product) format.
Gathering all components of $L_{i,\lambda}$ ($M_{i,\lambda}$) in corresponding core elements $L_i$ ($M_i$) in a row-wise (column-wise) fashion, see Appendix 2 of Ref.~\citep{Gelss2022a}, allows to express Hamiltonian $H$ in the following form
\begin{equation}
\begin{split}
    H & =  
    \left\llbracket\begin{matrix}
        S_1 & L_1 & I_1 & M_1 
    \end{matrix}\right\rrbracket
    \otimes 
    \left\llbracket\begin{matrix}
        I_2 & 0   & 0   & 0   \\
        M_2 & 0   & 0   & 0   \\       
        S_2 & L_2 & I_2 & 0   \\
        0   & 0   & 0   & J_2
    \end{matrix}\right\rrbracket
    \otimes \dots \\
		& \qquad \dots  \otimes
    \left\llbracket\begin{matrix}
        I_{N-1} & 0       & 0       & 0       \\
        M_{N-1} & 0       & 0       & 0       \\       
        S_{N-1} & L_{N-1} & I_{N-1} & 0       \\
        0       & 0       & 0       & J_{N-1}
    \end{matrix}\right\rrbracket
    \otimes
    \left\llbracket\begin{matrix}
        I_{N} \\
        M_{N} \\       
        S_{N} \\
        L_{N}
    \end{matrix}\right\rrbracket.
    \end{split}
		\label{eq:SLIM_2}
\end{equation}
where $J_i$ comprises $\xi_i$ identity matrices $I_i$ along the diagonal and zero matrices else.
Note that the double square bracket notation does not stand for block matrices but for the compact tensor notation of Ref.~\cite{Gelss2017}.
The Appendix of that work gives a proof of the above equation for all heterogeneous, cyclic systems. 
For homogeneous systems, the core elements $S_i$, $L_i$, $I_i$, $M_i$, and $J_i$ do not depend on the site index $i$. 

The ranks of the TT operator~\eqref{eq:SLIM_2} are naturally bounded due to the restriction to NN interactions only, e.g., for homogeneous and periodic systems, we have $\xi_1 = \dots = \xi_N =: \xi$ and, thus, $R = 2 + 2 \xi$, see~\cite{Gelss2017}.
One of the main advantages of SLIM decompositions is the linear scaling of the memory consumption with $N$ in case that the TT ranks of the solution do not increase with the order.
Similarly, this also holds for the computational effort when considering time-independent and -dependent Schrödinger equations, see Secs.~\ref{sec:dyn:tise} and \ref{sec:dyn:tdse}, respectively.
The considered SLIM decompositions in \Software are constructed using \textsc{Scikit-TT}.

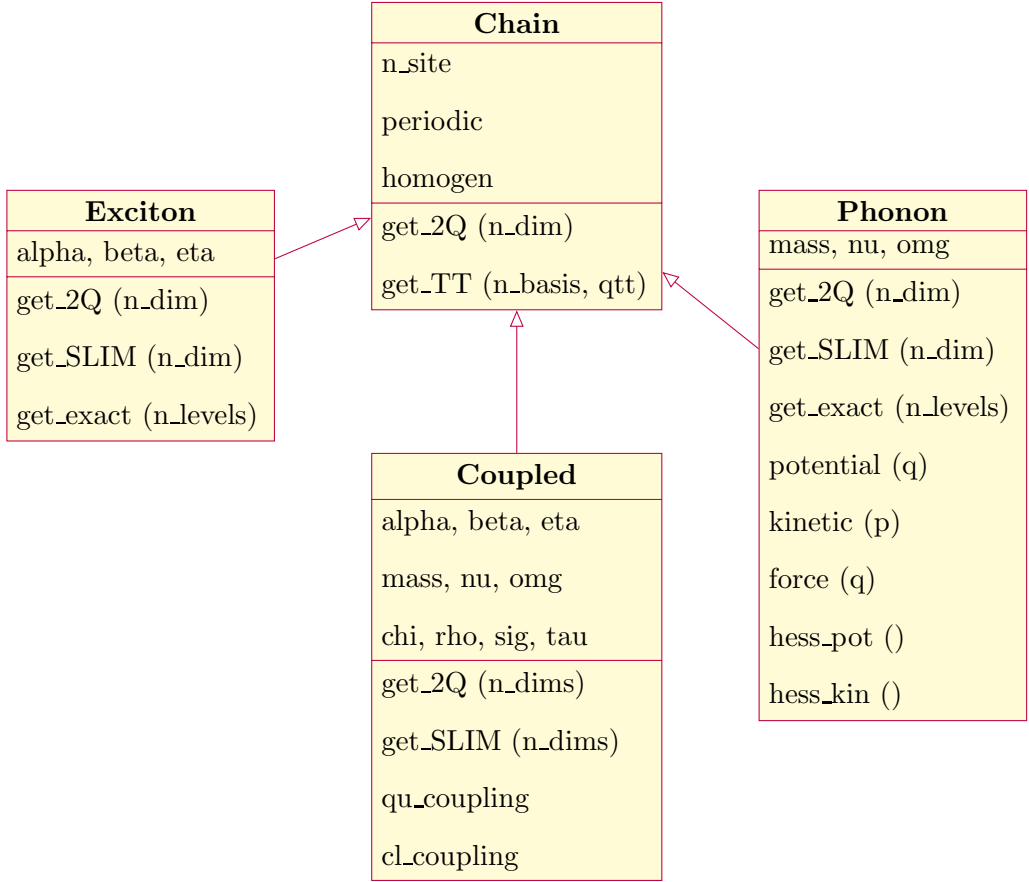
\begin{figure}[htbp]
\centering
\begin{tikzpicture}
    \begin{class}[text width=3.6cm]{Chain}{0,0} 
        \attribute{n\_site}
        \attribute{periodic}
        \attribute{homogen}
        \operation{get\_2Q (n\_dim)}
        \operation{get\_TT (n\_basis, qtt)}
    \end{class}

    \begin{class}[text width=3.3cm]{Exciton}{-5, -2.5}
        \inherit{Chain}
        \attribute{alpha, beta, eta}
        \operation{get\_2Q (n\_dim)}
        \operation{get\_SLIM (n\_dim)}
        \operation{get\_exact (n\_levels)}
    \end{class}

    \begin{class}[text width=3.3cm]{Phonon}{5, -2.5}
        \inherit{Chain}
        \attribute{mass, nu, omg}
        \operation{get\_2Q (n\_dim)}
        \operation{get\_SLIM (n\_dim)}
        \operation{get\_exact (n\_levels)}
        \operation{potential (q)}
        \operation{kinetic (p)}
        \operation{force (q)}
        \operation{hess\_pot ()}
        \operation{hess\_kin ()}
    \end{class}

    \begin{class}[text width=3.6cm]{Coupled}{0, -6}
        \inherit{Chain}
        \attribute{alpha, beta, eta}
        \attribute{mass, nu, omg}
        \attribute{chi, rho, sig, tau}
        \operation{get\_2Q (n\_dims)}
        \operation{get\_SLIM (n\_dims)}
        \operation{qu\_coupling}
        \operation{cl\_coupling}
    \end{class}

\end{tikzpicture}
\caption{Hierarchy of the Python classes representing the physical systems and Hamiltonians available as samples in \Software. 
Selected attributes and methods of each class are given in the upper and lower parts, respectively, of the boxes.
The corresponding Python files are located in folder wave\_train/hamilton.}
\label{fig:uml:ham}
\end{figure}

In the following subsections we will introduce exemplarily a few simple model Hamiltonians for chain-like systems with their SLIM decompositions and a description of the Python classes used for their respective implementations.
In particular, those are classes for excitons, for phonons, and for exciton-phonon coupling in quasi--1D chains.
Note that all these classes inherit from a common superclass for the implementation of the chain topology, see also the  
class hierarchy diagram shown in Fig.~\ref{fig:uml:ham}.

\subsection{Super class \textit{Chain}: General setup of linear or cyclic chain systems}
\label{sec:ham:chain}

The properties of the quasi-1D chain-like topologies underlying all of the present work are handled in super class \textit{Chain}.
For initialization, this class uses just three parameters. 
In addition to \texttt{n\_site} giving the number of sites, $N$ , the two Boolean variables \texttt{periodic} and \texttt{homogen} specify whether or not periodic boundary conditions are to be used and whether the chain is homogeneous or heterogeneous, respectively.
According to the latter setting, all further parameters of the respective Hamiltonians are given either as scalars or as Python lists.
Furthermore, the class \textit{Chain} contains two methods of general use: 
\paragraph{Method \textit{get\_2Q}} is intended for quantum-mechanical Hamiltonians formulated in terms of the second quantization.
For given dimension $d$ (argument \texttt{n\_dim}) of the local Hilbert space, which is assumed to be the same for each of the sites, this method sets up matrix representations of the raising ($a^\dagger$) and lowering ($a$) operators, as well as of the number operators.
Where applicable, the position and momentum operators are obtained from the ladder operators.

\paragraph{Method \textit{get\_TT}} is at the very heart of the SLIM formalism within our \Software package.
Given the lists of matrices $S_{i,\lambda},L_{i,\lambda},I_{i},M_{i,\lambda}$ (potentially independent of site index $i$ for homogeneous chains)  from one of the sub classes described below, this method serves to construct the tensor train super cores according to Eq.~\eqref{eq:SLIM_2}.
Subsequently, an instance of class \textit{TT} (tensor train) from the \textsc{Scikit-TT} package is created whose attributes (dimensions, ranks, and cores) are then set accordingly, depending on the initialization parameters \texttt{n\_site}, \texttt{periodic}, and \texttt{homogen}.

\subsection{Class \textit{Exciton}: Electronic Dynamics}
\label{sec:ham:exciton}

As a first example, we introduce a simple Hamiltonian for the excitonic dynamics of atoms or molecules in a chain-like arrangement. 
For simplicity, we restrict ourselves here to a chain of two-state-systems, e.g., assuming only excitations of one electron from the highest occupied to the lowest unoccupied molecular orbital (HOMO--LUMO).
Then, the excitonic Hamiltonian for a heterogeneous system of $N$ sites can be given in terms of (bosonic) exciton raising, $b_i^\dagger$, and lowering, $b_i$, operators for site $i$
\begin{equation}
	H^{\mathrm{(ex)}} = \sum_{i=1}^N \alpha_i  b_i^\dagger b_i
	                  + \sum_{i=1}^{N} \beta_i									
	                  \left(
	                  b_i^\dagger b_{i+1} + b_i b_{i+1}^\dagger
	                  \right)
                      + \eta
	\label{eq:H_ex}
\end{equation}
where the $\alpha_i$ are local ("on site") excitation energies and $\eta$ is a general offset of the energy scale.
The nearest-neighbor (NN) coupling energies $\beta_i$ between site $i$ and $i+1$, also known as "transfer integrals" or "hopping integrals", govern the delocalization and mobility of excitons within this simple model.
Here and throughout the following, the last summand ($i=N$) of the NN coupling term (with indices $i+1$ replaced by 1) is used for systems with periodic boundary conditions only and is omitted otherwise.

The most important methods in class \textit{Exciton} are described in the following:

\paragraph{Method \_\_init\_\_:}
The following sample input illustrates the handling of excitons within our \Software software package
{\small\begin{verbatim}
    from wave_train.hamilton.exciton import Exciton
    hamilton = Exciton(
        n_site=6, periodic=True, homogen=True, 
        alpha=0.1, beta=-0.01, eta=0.0
    )
\end{verbatim}}
This creates an object of class \textit{Exciton} the definition of which is imported from sub folder \textit{hamilton} in the \texttt{wave\_train} source folder.
Note that the first three arguments in the code above are used to initialize the super class \textit{Chain}, see Sec.~\ref{sec:ham:chain}, whereas the remaining three arguments specify the energetic parameters $\alpha,\beta,\eta$ as given in Eq.~\eqref{eq:H_ex}, the values of which are taken here from our previous work in Refs.~\citep{Gelss2022a,Gelss2022b}.

\paragraph{Method \textit{get\_SLIM}:} Based on the above attributes of class \textit{Exciton} and on the definition of the ladder operators in class \textit{Chain}, this method provides the SLIM formulation of Eq.~\eqref{eq:H_ex} yielding
\begin{eqnarray}
  S_i       = \alpha_i b_i^\dagger b_i + \frac{\eta}{N}I_i\nonumber \\
  L_{i,1}   = \beta_i b_i^\dagger,\quad M_{i+1,1} = b_{i+1}\nonumber \\
  L_{i,2}   = \beta_i b_i,\quad M_{i+1,2} = b_{i+1}^\dagger 
\label{eq:SLIM_ex}
\end{eqnarray}
where the dependence on the site index $i$ is omitted for the case of a homogeneous chain.
Note that method \textit{get\_SLIM} is called from within method \textit{get\_TT} in super class \textit{Chain} to construct the tensor train supercores according to Eq.~\eqref{eq:SLIM_2}, see Sec.~\ref{sec:ham:slim}.
The following code line illustrates this for the case of excitons
{\small\begin{verbatim}
    hamilton.get_TT(n_basis=2, qtt=False)
\end{verbatim}}
where the first argument gives the dimension $d$ of the local exciton Hilbert space, i.e., the size of the electronic basis set.

\paragraph{Method \textit{get\_exact}:} 
For the case of homogeneous excitonic chains, i.e., with all sites being equivalent, this method provides analytic/exact solutions of the time-independent Schr\"{o}dinger equation (TISE) based on a Bethe ansatz as given in Ref.~\citep{Gelss2022a}.
In principle, the number of analytic solutions to be calculated can be chosen by the user, see Sec.~\ref{sec:dyn:tise} below.
However, for linear systems, only the energy levels for the ground state and for the $N$ states within the Fock space of singly excited states are currently available, which are obtained in close analogy to H\"{u}ckel theory.
For cyclic systems, we also implemented the $N(N-1)/2$ energy levels for states with two quanta of excitation~\citep{Hu2018}.

\subsection{Class \textit{Phonon}: Vibrational Dynamics}
\label{sec:ham:phonon}

As another example, we introduce a simple Hamiltonian for the vibrational (phononic) dynamics of a one-dimensional lattice model based on the harmonic approximation.
In terms of site masses $m_i$, displacement coordinates $R_i$, and conjugate momenta $P_i$, a general Hamiltonian can be written as 
\begin{equation}
	H^{\mathrm{(ph)}} 
	= \frac{1}{2} \sum_{i=1}^N \frac{P_i^2}{m_i}
	+ \frac{1}{2} \sum_{i=1}^N m_i \nu_i^2 R_i^2
	+ \frac{1}{2} \sum_{i=1}^N \mu_i \omega_i^2 
	\left( R_i-R_{i+1} \right)^2
	\label{eq:H_ph1}
\end{equation}
where each site $i$ is restrained around its equilibrium position by harmonic oscillators with frequencies $\nu_i$.
The NN interactions between neighboring sites $i$ and $i+1$ are modeled by harmonic oscillators with frequency $\omega_i$ and corresponding reduced masses $\mu_i=m_im_{i+1}/(m_i+m_{i+1})$.

In analogy to the treatment of the excitons in Sec.~\ref{sec:ham:exciton}, we re-formulate the phononic Hamiltonian of Eq.~(\ref{eq:H_ph1}) using second quantization 
\begin{equation}
	H^{\mathrm{(ph)}} = \sum_{i=1}^N \tilde{\nu}_i
	                    \left(c_i^\dagger c_i + \frac{1}{2}\right)
	                  - \sum_{i=1}^N \tilde{\omega}_i
					\left( c_i^\dagger+c_i \right)
					\left( c_{i+1}^\dagger+c_{i+1} \right)
	\label{eq:H_ph2}
\end{equation}
with raising ($c_i^\dagger$) and lowering ($c_i$) operators of (local) vibrations of site $i$.
The effective frequencies of single site and NN pair vibrations are given by
\begin{eqnarray}
\tilde{\nu}_i&=&\sqrt{\nu_i^2+\frac{m_{i-1}}{m_i+m_{i-1}}\omega_{i-1}^2+\frac{m_{i+1}}{m_i+m_{i+1}}\omega_{i}^2} \label{eq:nu_E} \\
\tilde{\omega}_i&=&\frac{\mu_i\omega_{i}^2}{2\sqrt{m_i \tilde{\nu}_i m_{i+1} \tilde{\nu}_{i+1}}}
\label{eq:omg_E}
\end{eqnarray}
where for linear systems without periodic boundary conditions the second or third term under the square root of Eq.~\eqref{eq:nu_E} are omitted for the first ($i=1$) or last ($i=N$) site, respectively.
Note that the SLIM structure defined in Eq.~\eqref{eq:SLIM_1} is apparent in our formulation~\eqref{eq:H_ph2} for the phononic Hamiltonian.

The most important methods in class \textit{Phonon} are described in the following:

\paragraph{Method \_\_init\_\_:}
This sample input shows the setup of the phonon dynamics using the \Software package
{\small\begin{verbatim}
    from wave_train.hamilton.phonon import Phonon
    hamilton = Phonon(
        n_site=6, periodic=True, homogen=True, 
        mass=1, nu=1e-3, omg=2**(1/2)1e-3
    )
\end{verbatim}}
which creates an object of class \textit{Phonon} which is imported from subfolder \textit{hamilton} in the \texttt{wave\_train} source folder.
Again, the first three arguments in the code above are used to initialize the super class \textit{Chain}, see Sec.~\ref{sec:ham:chain}, whereas the other three arguments specify the masses and frequency parameters $m,\nu,\omega$ as given in Eq.~\eqref{eq:H_ph1}.
The initialization method of class \textit{Phonon} also provides the effective frequencies $
\tilde{\nu}$ and $\tilde{\omega}$, see Eqs.~\eqref{eq:nu_E},~\eqref{eq:omg_E}.

\paragraph{Method \textit{get\_SLIM}:} 
Based on the above attributes of class \textit{Phonon} and on the matrix representations of the ladder operators from super class \textit{Chain}, the SLIM formulation of Eq.~\eqref{eq:H_ph2} is straight-forwardly expressed as 
\begin{eqnarray}
 S_i       =  \tilde{\nu}_i \left(c_i^\dagger c_i + \frac{1}{2}\right) \nonumber \\
 L_{i,1}   = - \tilde{\omega}_i\left( c_i^\dagger+c_i \right),\quad
 M_{i+1,1} = c_{i+1}^\dagger+c_{i+1}
\label{eq:SLIM_ph}
\end{eqnarray}
where the dependence on the site index $i$ becomes irrelevant for a homogeneous chain.
Again, method \textit{get\_SLIM} is called from within method \textit{get\_TT} (super class \textit{Chain}), to construct the tensor train supercores, see Eq.~\eqref{eq:SLIM_2}.
The use of this method is illustrated here
{\small\begin{verbatim}
    hamilton.get_TT(n_basis=8, qtt=False)
\end{verbatim}}
where the first argument gives the dimension $d$ of the local phonon Hilbert space, i.e., the size of the harmonic oscillator vibrational basis set.
In practice, this parameter needs to be determined by convergence tests.
Typically, it depends on the total energy available in the simulated system.
\paragraph{Method \textit{get\_exact}:}
Also for the one-dimensional chain of oscillators given in Eqs.~\eqref{eq:H_ph1}, we implemented reference solutions for homogeneous chains to check the accuracy of the numeric TISE solvers described in Sec.~\ref{sec:dyn:tise} below.
For periodic chains, analytic (Bloch type) solutions are well known, see our previous work~\citep{Gelss2022a}.
For non-periodic systems, where fully analytic solutions are not available because of the non-uniformity of the effective frequencies in Eqs.~\eqref{eq:nu_E} and~\eqref{eq:omg_E}, energy levels are obtained from a conventional normal mode analysis which is considered to be quasi-exact here. 
Note that this requires the calculation of the Hessian matrix of the phonon potential energy function of Eq.~\eqref{eq:H_ph1} which is provided in method \textit{hess\_pot} in class \textit{Phonon}, see also Fig.~\ref{fig:uml:ham}.

\subsection{Class \textit{Coupled}: Exciton-Phonon-Coupling}
\label{sec:ham:coupled}

Because the excitonic energy transfer is known to be affected by coupling to vibrational degrees of freedom, the study of exciton-phonon coupling (EPC) is of vital importance, e.g. for the transport of electronic energy in semiconducting materials \citep{Mikhnenko2015,Schroter2015,Zhugayevych2015} or the transport of amide I vibrational energy in helical proteins \cite{Scott1991, Georgiev2019}.
Within the Hilbert space used for EPC, which is a direct product of the Hilbert spaces for the excitonic and phononic states, the total Hamiltonian can be written as
\begin{equation}
	H = H^{\mathrm{(ex)}} \otimes \mathbb{I}^{\mathrm{(ph)}} 
	+ \mathbb{I}^{\mathrm{(ex)}} \otimes H^{\mathrm{(ph)}} 
	+ H^{\mathrm{(epc)}} 
	\label{eq:H_total}
\end{equation}
where $H^{\mathrm{(ex)}}$ and $H^{\mathrm{(ph)}}$ are the Hamiltonians for excitons and phonons, see Eqs.~\eqref{eq:H_ex} and \eqref{eq:H_ph2}, and where $\mathbb{I}^{\mathrm{(ex)}}$ and $\mathbb{I}^{\mathrm{(ph)}}$ are identity operators on the respective Hilbert spaces.
A selection of simple, Fr\"{o}hlich-Holstein type Hamiltonians $H^\mathrm{(epc)}$ for the coupling of excitons and phonons is implemented in \Software
\begin{eqnarray}
	\sum_{i=1}^N \chi_i  b_i^\dagger b_i \otimes R_i &=&
    \sum_{i=1}^N \bar{\chi}_i b_i^\dagger b_i \otimes 
      \left( c_i^\dagger+c_i \right) \nonumber \\
    \sum_{i=1}^N  \rho_i  b_i^\dagger b_i \otimes (R_{i+1}-R_i) &=&          
	\sum_{i=1}^N  b_i^\dagger b_i \otimes \left[
		\bar{\rho}_i       \left(c_{i+1}^\dagger+c_{i+1} \right) -
		\bar{\bar{\rho}}_i \left(c_i^\dagger+c_i\right)
		\right] \nonumber \\
	\sum_{i=1}^N \sigma_i b_i^\dagger b_i \otimes 
		\left( R_{i+1} - R_{i-1} \right) &=& 
	\sum_{i=1}^N 
	    b_i^\dagger b_i \otimes \left[
		\bar{     \sigma}_i	\left( c_{i+1}^\dagger+c_{i+1} \right) -
		\bar{\bar{\sigma}}_i	\left( c_{i-1}^\dagger+c_{i-1} \right)
		\right] \nonumber \\
	\sum_{i=1}^N \tau_i  \left(b_i^\dagger b_{i+1} +b_i b_{i+1}^\dagger \right)	
		\otimes \left(R_{i+1} - R_i \right)&=& 
	\sum_{i=1}^N \left(b_i^\dagger b_{i+1}+b_i b_{i+1}^\dagger \right) \otimes 			\left[
		\bar{\tau}_i       \left( c_{i+1}^\dagger+c_{i+1} \right) -
		\bar{\bar{\tau}}_i \left( c_i^\dagger+c_i \right) \right] \nonumber \\
	\label{eq:H_epc}
\end{eqnarray}
Here the EPC constants $\chi$, $\rho$, and $\sigma$ give the linear dependence of the excitonic site energies $\alpha$ on the positions of, or distances between, nearest or second-nearest sites, respectively.
In contrast, the constants $\tau$ characterize the dependence of excitonic coupling energies $\beta$ on the corresponding distances thus including also Holstein-Peierls type models.
The bar notation in Eq.~\eqref{eq:H_epc} is used to convert the EPC constants to second quantization
\begin{eqnarray}
\bar{\chi}_i=\chi_i/\sqrt{2m_{i}\tilde{\nu}_{i}}, \nonumber \\
\bar{\rho}_i=\rho_i/\sqrt{2m_{i+1}\tilde{\nu}_{i+1}},&\quad&
\bar{\bar{\rho}}_i=\rho_i/\sqrt{2m_i\tilde{\nu}_i},\nonumber \\
\bar{\sigma}_i=\sigma_i/\sqrt{2m_{i+1}\tilde{\nu}_{i+1}},&\quad& \bar{\bar{\sigma}}_i=\sigma_i/\sqrt{2m_{i-1}\tilde{\nu}_{i-1}} \nonumber \\
\bar{\tau}_i=\tau_i/\sqrt{2m_{i+1}\tilde{\nu}_{i+1}}&,\quad&
\bar{\bar{\tau}}_i=\tau_i/\sqrt{2m_i\tilde{\nu}_i}
\end{eqnarray}
Note that in our previous work~\cite{Gelss2022a}, the distinction between EPC constants with bars and double bars was missing, which, however, was not required for the cyclic systems mainly investigated there.

In the following, a description of important methods comprising class \textit{Coupled} will be given:

\paragraph{Method \_\_init\_\_:}
The following lines of input serve to create an instance of class \textit{Coupled}
{\small\begin{verbatim}
    from wave_train.hamilton.coupled import Coupled    
    hamilton = Coupled(
        n_site=5, periodic=True, homogen=True, 
        alpha=0.1, beta=-0.01, eta=0.0,
        mass=1, nu=1e-3, omg=1e-3*2**(1/2),
        chi=0, rho=0, sig=1.6e-4, tau=0
    )
\end{verbatim}}
where the first nine arguments specify the chain topology, the excitons, and the phonons, see Secs.~\ref{sec:ham:chain}, \ref{sec:ham:exciton}, \ref{sec:ham:phonon}, respectively. 
The last four arguments specify the parameters $\chi,\rho,\sigma,\tau$ required for the different types of EPC models given in Eq.~\eqref{eq:H_epc}.
For simplicity, we only consider the $\sigma$--coupling mechanisms in the present work.

\paragraph{Method get\_2Q:} Unlike classes \textit{Exciton} and \textit{Phonon}, which essentially use the inherited  method \textit{get\_2Q} from super class \textit{Chain}, the class \textit{Coupled} overrides the super class method \textit{get\_2Q}.
Here, one object of class \textit{Exciton} and another object of class \textit{Phonon} are created, along with their respective matrix representations for ladder operators.
This allows a convenient calculation of direct products of excitonic and phononic operators, e.g., $b_i \otimes c_{i+1}$, using the Numpy function \texttt{kron} for the Kronecker product.

\paragraph{Method get\_SLIM:} This method is intended to provide the SLIM formulation of Eq.~\eqref{eq:H_epc} which is given by
\begin{eqnarray}
    && S_i = ( \bar{\chi}_i - \bar{\bar{\rho}}_i ) 
	b_i^\dagger b_i \otimes \left( c_i^\dagger+c_i \right) \nonumber \\
 L_{i,1}   = (\bar{\rho}_i+\bar{\sigma}_i) b_i^\dagger b_i &, \quad & 
 M_{i+1,1} = c_{i+1}^\dagger+c_{i+1} \nonumber \\
 L_{i,2}   = - \left( c_i^\dagger+c_i \right) &, \quad & 
 M_{i+1,2} = \bar{\bar{\sigma}}_{i+1} b_{i+1}^\dagger b_{i+1} \nonumber \\
 L_{i,3}   = \bar{\tau}_i b_i^\dagger &,\quad & 
 M_{i+1,3} = b_{i+1}\otimes\left( c_{i+1}^\dagger+c_{i+1} \right), \nonumber \\
 L_{i,4}   = -\bar{\bar{\tau}}_i b_i^\dagger\otimes\left( c_i^\dagger+c_i \right) &,\quad & 
 M_{i+1,4} = b_{i+1}, \nonumber \\
 L_{i,5}   = \bar{\tau}_i b_i &,\quad & 
 M_{i+1,5} = b_{i+1}^\dagger\otimes\left( c_{i+1}^\dagger+c_{i+1} \right), \nonumber \\
 L_{i,6}   = -\bar{\bar{\tau}}_i b_i\otimes\left( c_i^\dagger+c_i \right) &,\quad & 
 M_{i+1,6} = b_{i+1}^\dagger
\label{eq:SLIM_epc}
\end{eqnarray}
Also here, the method \textit{get\_SLIM} is called within method \textit{get\_TT} of super class \textit{Chain} which constructs the tensor train super cores, see Sec.~\ref{sec:ham:chain}.
The following code line illustrates this for the case of coupled excitons and phonons
{\small\begin{verbatim}
    hamilton.get_TT(n_basis=[2, 8], qtt=False)
\end{verbatim}}
where the Python list in the first argument contains the sizes of the electronic and vibrational basis sets, respectively.

\section{Quantum and Classical Dynamics}
\label{sec:dyn}
\subsection{Super classes for quantum and classical mechanics}
\label{sec:dyn:super} 

This section deals with the implementation of different types of physical/chemical dynamics within \Software.
The main work horses of our software package are  the classes \textit{TISE} and \textit{TDSE} containing numerical solvers for the time-independent and time-dependent Schr\"{o}dinger equation based on the TT tensor format, see Secs.~\ref{sec:dyn:tise} and \ref{sec:dyn:tdse}.
For completeness, we have added classes \textit{QCMD} and \textit{CEoM} for mixed quantum-classical molecular dynamics and fully classical dynamics, see Secs.~\ref{sec:dyn:qcmd} and \ref{sec:dyn:ceom}.

The four main classes inherit from a set of super classes for quantum mechanics, mixed quantum-classical mechanics, and classical mechanics, see Fig.~\ref{fig:uml:dyn} for a class hierarchy diagram.
Upon initializing objects of any of these classes, an input argument \texttt{hamilton} is required, which has to be an object of one of the three classes for excitons, phonons, or coupled systems explained above in Sec.~\ref{sec:ham}.
Note that quantum-classical dynamics only works for coupled exciton--phonon systems while fully classical dynamics is restricted to phonons only.
Most importantly, each of the three super classes provides a method \textit{observe} which deals with calculating and printing expectation values of important observables such as energy, positions and momenta of the particles.
This is complemented by utility methods such as calculations of "braket" scalar products, expectation values with their uncertainties, and reduced density matrices for quantum simulations.

In turn, the three super classes inherit from the more fundamental class \textit{Mechanics} for general mechanical systems.
This class contains method \textit{save} which writes important quantities into binary data files which can be either of Python '\textit{pickle}' or of Matlab '\textit{mat}' type.
Those file types can also be read by method \textit{load} which thus serves to obtain deviations between the results of two simulations, e.g., for the case of different dynamic or different numerical schemes applied to the same physical problem and the same time discretization.
Note that such a comparison is based on root mean squared deviations (RMSD), either for the quantum state vectors themselves, for populations, or for expectation values of observables such as positions or momenta.
The corresponding file names for such a comparison are set as properties \texttt{save\_file} and \texttt{load\_file}, and the type of comparison is set by the string \texttt{compare}.
Moreover, class \textit{Mechanics} also contains a method for linear regressions of conserved quantities, such as energy or norm of state vectors, or of the mentioned RMSDs.
Finally, methods \texttt{gaussian} and \texttt{sec\_hyp} can be used to set up  wave packets with Gaussian or hyperbolic secant envelope, respectively, see the description of the eligible sub classes in Secs.~\ref{sec:dyn:tdse}, \ref{sec:dyn:qcmd}.
\begin{turnpage}
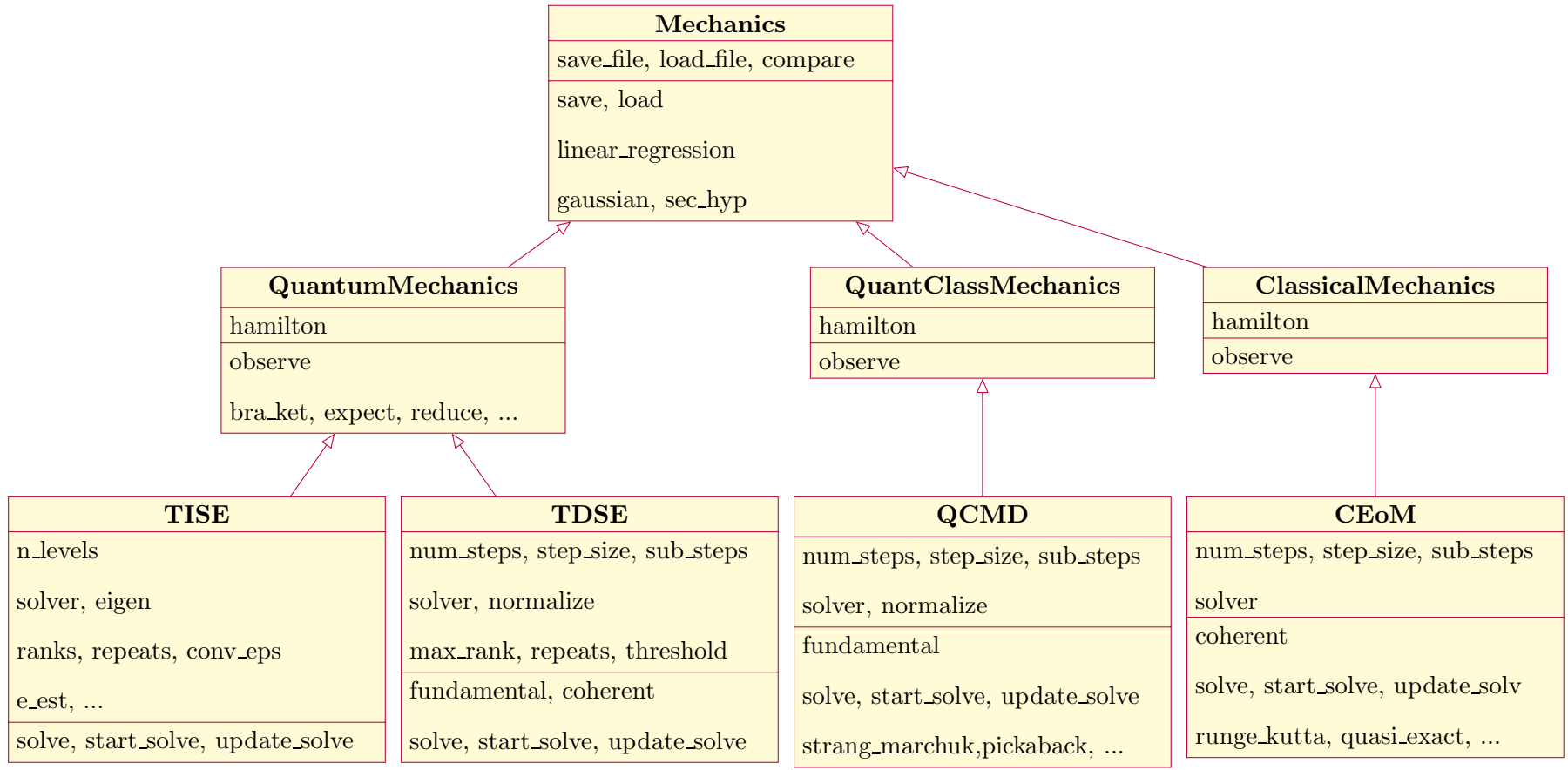
\begin{figure}[htbp]
\centering
\begin{tikzpicture}
    \begin{class}[text width=5cm]{Mechanics}{10,0} 
    		\attribute{save\_file, load\_file, compare}
        \operation{save, load}
        \operation{linear\_regression}
        \operation{gaussian, sec\_hyp}
    \end{class}
    
    \begin{class}[text width=5cm]{QuantumMechanics}{5, -4} 
        \inherit{Mechanics}
        \attribute{hamilton}
        \operation{observe}
        \operation{bra\_ket, expect, reduce, ...} 
    \end{class}
    
    \begin{class}[text width=5cm]{QuantClassMechanics}{14, -4} 
        \inherit{Mechanics}
        \attribute{hamilton}
        \operation{observe}
    \end{class}

    \begin{class}[text width=5cm]{ClassicalMechanics}{20, -4} 
        \inherit{Mechanics}
        \attribute{hamilton}
        \operation{observe}
    \end{class}
    
    \begin{class}[text width=5.5cm]{TISE}{2, -7.5}
        \inherit{QuantumMechanics}
        \attribute{n\_levels}
        \attribute{solver, eigen}
        \attribute{ranks, repeats, conv\_eps}
        \attribute{e\_est, ...}
        \operation{solve, start\_solve, update\_solve}
    \end{class}

    \begin{class}[text width=5.5cm]{TDSE}{8, -7.5}
        \inherit{QuantumMechanics}
        \attribute{num\_steps, step\_size, sub\_steps}
        \attribute{solver, normalize}
        \attribute{max\_rank, repeats, threshold}
        \operation{fundamental, coherent}
        \operation{solve, start\_solve, update\_solve}
    \end{class}

    \begin{class}[text width=5.5cm]{QCMD}{14, -7.5}
        \inherit{QuantClassMechanics}
        \attribute{num\_steps, step\_size, sub\_steps}
        \attribute{solver, normalize}
        \operation{fundamental}
        \operation{solve, start\_solve, update\_solve}
        \operation{strang\_marchuk,pickaback, ...}
    \end{class}

    \begin{class}[text width=5.5cm]{CEoM}{20, -7.5}
        \inherit{ClassicalMechanics}
        \attribute{num\_steps, step\_size, sub\_steps}
        \attribute{solver}
        \operation{coherent}
        \operation{solve, start\_solve, update\_solv}
        \operation{runge\_kutta, quasi\_exact, ...}
    \end{class}

\end{tikzpicture}
\caption{Hierarchy of the Python classes for quantum and classical dynamics available in \Software. 
Selected attributes and methods of each class are given in the upper and lower parts, respectively, of the corresponding boxes.
The corresponding Python files are located in folder wave\_train/dynamics.}
\label{fig:uml:dyn}
\end{figure}
\end{turnpage}

\subsection{Class \textit{TISE}: Time-Independent Schr\"{o}dinger Equation}
\label{sec:dyn:tise}

Solving the time-independent Schr\"{o}dinger equation (TISE) 
\begin{equation}
\hat{H} |\Psi_n\rangle = E_n |\Psi_n\rangle, \quad n=0, 1, \ldots
\label{eq:tise}
\end{equation}
yields a set of stationary quantum states $|\Psi_n\rangle$ along with their corresponding energies $E_n$ where $\hat{H}$ is one of the (time-independent) Hamiltonians presented in Sec.~\ref{sec:ham} or another one provided by the  user.
To beat the curse of dimensionality, the strategy followed in the \Software software builds on low-rank tensor approximations for the state vectors, in analogy to the TT representation of the Hamiltonian given in Sec.~\ref{sec:ham:slim}.
In practice, the eigenvalue problem is solved numerically using the alternating linear scheme (ALS) which is an iterative algorithm based on sequential contractions of the TT cores of $\hat{H}$ and $|\Psi_n\rangle$ to construct low-dimensional eigenvalue problems~\cite{Holtz2012}.
A key feature of the \Software implementation is that not only ground states but also higher excited states can be obtained in an efficient way by means of integrated Wielandt deflation which enables us to displace
previously computed eigenvalues while keeping all other eigenvalues unchanged, see~\citep{Gelss2022a}.
To avoid an explosion of the computational costs for higher excited states, which would arise in a straight-forward application of the Wielandt deflation, the computation of the deflated Hamiltonians is implicitly incorporated into the ALS routine of \textsc{Scikit-TT}.

The use of class \textit{TISE} is illustrated here by the following sample input 
{\small\begin{verbatim}
   from wave_train.dynamics.tise import TISE
   dynamics = TISE(hamilton=hamilton, n_levels=10,
                   solver='als', eigen='eigs',
                   ranks=15, repeats=20, conv_eps=1e-8,
                   e_est=0.08)
   dynamics.solve()
\end{verbatim}}
where object \texttt{hamilton} pertains to one of the classes described in Sec.~\ref{sec:ham} and \texttt{n\_levels} gives the desired number of eigenvalues to be calculated.
The argument \texttt{solver} serves to choose the scheme to solve the full eigenproblem, by default the above-mentioned ALS algorithm which is one of the key components of the \texttt{scikit\_tt} package.
The next argument, \texttt{eigen}, specifies the solver used for the micro-problems within each of the ALS iterations, in this case the sparse matrix eigensolver \texttt{'eigs'} from the SciPy package.
Alternative choices are \texttt{'eig'} or \texttt{'eigh'}.
The subsequent arguments serve to specify the  ALS parameters, most importantly the number (\texttt{ranks}) of maximal ranks of the solutions.
In all cases, ALS iterations are terminated once the estimated eigenvalues do not change by more than a certain threshold (\texttt{conv\_eps}) in the last three ALS sweeps or when the number of sweeps reaches the limit given by attribute \texttt{repeats}.
Finally, the parameter \texttt{e\_est} gives an estimated energy (here: $\alpha-2|\beta|$) close to which the energy levels are to be searched.
If \texttt{eigen} is set to \texttt{'eig'}, eigenvalues closest to \texttt{e\_est} are chosen from the list of all computed eigenvalues. Otherwise Scipy's \texttt{'eigs'} uses the shift-invert mode to find the desired eigenvalues.
This is of importance, e.g., when calculating the stabilization gained from mutual trapping of phonons and excitons from the lowest eigenvalue within the $\mathcal{N}^\mathrm{(ex)}=1$ manifold~\citep{Gelss2022a}.
Typically, below that energy there is a huge number of eigenvalues in the $\mathcal{N}^\mathrm{(ex)}=0$ manifold which are not of interest and which can thus be excluded.

As an alternative, the \Software package offers quasi-exact solutions, provided that the dimension of the full Hibert space, $d^N$, is not too large (typically 4096 for a standard PC).
In that case, tensor train methods are bypassed and the eigenproblem for a matricized version of $H$ is solved directly. This is invoked by setting \texttt{solver = 'qe'} where the parameter \texttt{eigen} again specifies the choice of the numeric solver.
While this option is clearly not eligible for longer chains, it serves the purpose of creating reference solutions for shorter chains.

The resulting energy levels will also be compared against analytic (or semi-analytic) solutions which are available only for the Hamiltonians \eqref{eq:H_ex} for uncoupled excitons and \eqref{eq:H_ph2} for uncoupled phonons, see also \cite{Gelss2022a}.

\subsection{Class \textit{TDSE}: Time-Dependent Schr\"{o}dinger Equation}
\label{sec:dyn:tdse}

The evolution of quantum states, $\Psi(t)$, is obtained as a solution of the time-dependent Schr\"{o}dinger equation (TDSE) for one of the  Hamiltonians $\hat{H}$ of Sec.~\ref{sec:ham}
\begin{equation}
\mathrm{i} \frac{\mathrm d}{\mathrm dt} |\Psi(t)\rangle = \hat{H} |\Psi(t)\rangle,\quad |\Psi(t=0)\rangle = |\Psi_0\rangle
\label{eq:tdse}
\end{equation}
where atomic units with $\hbar=1$ are used.
Again, the problem of high dimensionality is tackled by  strategies building on low-rank tensor representations of the state vectors.
Our implementation of class \textit{TDSE} within \Software builds on the choice of numeric propagators for tensor trains available within the \texttt{scikit\_tt} software package. 
Restricting ourselves to explicit, reversible, and symplectic schemes, the most obvious choice is a symmetric, second order Euler (S2) method.
This method has been routinely used in the quantum dynamics community for several decades, where it is also known as second-order differencing scheme \citep{Askar1978,Leforestier1991}.
Within \Software, also higher order variants, e.g. fourth (S4) and sixth (S6) order differencing methods are available.
The former one has been shown to offer a good compromise between efficiency and accuracy~\citep{Gelss2022b}.

Frequently used alternatives are based on operator splitting originally developed for cases where Hamiltonians consist of kinetic and potential energy which are treated separately in momentum and position representation, respectively \citep{Fleck1976,Feit1982}.
In the present work, however, we resort to the  Hamiltonians of Sec.~\ref{sec:ham:chain} for systems with a chain-like topology and NN interactions only.
For such cases, various novel splitting schemes are available in \textsc{Scikit-TT} which are based on separating the interlacing pairs of NN sites~\citep{Orus2014,Paeckel2019,Gelss2022b}.
Not only the classical first order Lie-Trotter (LT) and second order Strang-Marchuk (SM) schemes, but also higher-order compositions of the basic methods are available, namely the 4-th order Yoshida-Neri (YN) and the 8-th order Kahane-Li (KL) method which have displayed an excellent accuracy in our test calculations~\citep{Gelss2022b}.
For more information see Ref.~\cite{Lubich2008}, where an overview of splitting methods with different order is given.

Finally, note that implicit schemes such as the trapezoidal rule or the midpoint rule are available within \texttt{scikit\_tt}, too. 
However, in our quantum dynamics test simulations they have displayed a very unfavorable numeric effort because they involve the solutions of large-scale linear systems of equations.
While the use of ALS~\cite{Holtz2012} is an integral part of the \textit{TISE} class, doing this at each time step results in an unfavorable numerical effort.
Therefore, the \textit{TDSE} class is solely based on explicit integration schemes.

To demonstrate the use of the \textit{TDSE} class we consider the following sample code lines
{\small\begin{verbatim}
   from wave_train.dynamics.tdse import TDSE
   dynamics = TDSE(hamilton=hamilton, 
       num_steps=50, step_size=20, sub_steps=5,
       solver='s2', normalize=0,
       max_rank=8, threshold=1e-12) 
\end{verbatim}}
where the object \texttt{hamilton} refers to one of the Hamiltonian classes of Sec.~\ref{sec:ham}.
Here we propagate for 1000 (atomic) units of time, divided into 50 main time steps with a (constant) length of 20 units.
After each of the main steps, expectation values of important observables are calculated and printed, and a frame is added to the (optionally generated) animated visualization, see Sec.~\ref{sec:gra}.
Internally, each of the main steps can be divided into a (constant) number of sub steps (here 5).
The arguments \texttt{solver} and \texttt{normalize} for the initialization of class \textit{TDSE} specify the choice of the numeric solver (two-letter codes explained above) as well as whether normalization of the state vector after every sub step is to be enforced or not.
The remaining arguments are \texttt{max\_rank}, the maximal rank in the decomposition of solutions $\Psi(t)$, and \texttt{threshold} the value of which is used for the rank truncation within the splitting schemes (LT, SM, YN, KL) and the symmetric Euler (S2, S4, S6) schemes.
In both cases, an orthonormalization scheme called \emph{higher-order singular value decomposition} (HOSVD)~\cite{Oseledets2011} with absolute as well as relative cut-off criteria for singular values is applied to keep the TT ranks of our solutions bouned by \texttt{max\_rank}.
 
Before actually solving the TDSE, it is necessary to specify the initial state $|\Psi(t=0)\rangle = |\Psi_0\rangle$. 
To that end, the class \textit{TDSE} contains method \texttt{fundamental} to set up an initial state where one (or more) sites are fundamentally ($0 \rightarrow 1$) excited while all others are prepared in their ground state.
The resulting quantum state is constructed as a tensor train using the \texttt{TT} class from the \textsc{Scikit-TT} toolbox. 
That is, depending on a given vector of coefficients \texttt{coeffs}, the canonical representation of $|\Psi_0\rangle$ is given by the sum over tensor products of the form \texttt{coeff}$[j] \cdot \bigotimes_{k=1}^n v_j^{(k)}$ for non-zero coefficients, where $v_j^{(k)} = [0,1]^\top$ if $k=j$ and otherwise $[1,0]^\top$.
The created instance of the  \texttt{TT} class then stores the cores of the corresponding TT representation of $|\Psi_0\rangle$

While this method works in an analogous way for excitons and phonons, we note that for coupled systems only the electronic parts are fundamentally excited whereas the vibrational parts are in their ground states. 
In the following Python example
{\small\begin{verbatim}
   dynamics.fundamental()
   dynamics.solve()
\end{verbatim}}
the default behavior is to return a state with a single excitation localized at the central site of the chain, which then serves as an initial state for solving the TDSE.
It is also possible to give a vector of coefficients as input for method \textit{fundamental}, in which case a weighted sum of products, each with a single site excitation, is returned. 
This feature of \Software can be used, e.g., to construct bell-shaped wave packets with Gaussian or hyperbolic secant (sech) envelope with settable mean position, mean momentum, and width.
The Gaussian shape is typically used to describe a free particle whereas the  sech shape typically occurs as a solution of nonlinear cubic TDSEs, see e.g. Davydov's soliton theory \citep{Davydov1985,Georgiev2019}.

As an alternative to the use of fundamentally excited states, class \textit{TDSE} also contains method \texttt{coherent} which is meant only for vibrational systems, see our description of class \textit{Phonons} in Sec.~\ref{sec:ham:phonon}.
That method serves to set up coherent states of the $i$--th site which are eigenstates of the lowering operator $c_i$, defined as $c_i|\zeta\rangle_i = \zeta_i |\zeta\rangle_i$   with
\begin{equation}
|\zeta\rangle_i = \mathrm{e}^{- \frac{|\zeta_i|^2}{2}} \sum_{k=0}^\infty \frac{\zeta_i^k}{\sqrt{k!}} |k\rangle_i
\label{eq:coherent}
\end{equation} 
Here $|k\rangle_i$ stands for the $k$-th harmonic oscillator eigenstate of the $i$-th site and
\begin{equation}
\langle R_i \rangle = \sqrt{\frac{2}{m_i\tilde{\nu}_i}}\zeta_i
\label{eq:displace}
\end{equation}
gives the mean value of the displacement coordinate, $R_i$, of the respective quantum harmonic oscillator with mass $m_i$ and effective frequency $\tilde{\nu}_i$.
In analogy to method \texttt{fundamental}, also method \texttt{coherent}  allows for the possibility of a combination of excitations of several sites.

In close analogy to class \textit{TISE} described in Sec.~\ref{sec:dyn:tise}, also class \textit{TDSE} offers quasi-exact solutions for simulations where the full Hilbert space dimension is not too large.
In that case, the matricized Hamiltonian is exponentiated yielding a direct way to calculate the time evolution operator.
This can be useful when benchmarking the accuracy of different propagation schemes and/or different time steps, see e.g. our results in Ref.~\citep{Gelss2022b}. 
Moreover, for two-state systems, class \textit{TDSE} calculates analytic Bessel function solutions of the time evolution~\citep{Kenkre1984}, e.g., for class \textit{Exciton} explained in Sec.~\ref{sec:ham:exciton}.
However, their use for benchmarking TT-based solutions is limited because they build on the assumption of non-periodic, infinitely-long chains.

\subsection{Class \textit{QCMD}: Quantum-Classical Molecular Dynamics}
\label{sec:dyn:qcmd}

The above-mentioned TT-based approaches implemented in \Software can be very helpful instruments in tackling problems in quantum dynamics of bipartite systems such as the example of coupled excitons and phonons mentioned in Sec.~\ref{sec:ham:coupled}.
On the one hand, we have shown that the computational effort is almost linear in $N$ which allows for treating long chains~\citep{Gelss2022a,Gelss2022b}.
On the other hand, these methods can mitigate the \textit{curse of dimensionality} only as long as the problem at hand allows for an acceptable accuracy of the approximate solution when we restrict ourselves to TT cores with ranks of manageable size.
However, the computational effort for solving the TDSE scales at least with  $d^2$ (symmetric Euler) where $d$ is the dimension of the local Hilbert space.
Hence, there are still simulation scenarios where a fully quantum-mechanical treatment is out of reach with the computational resources of today, and probably also in the foreseeable future.

In many simulation scenarios, a clear separation of time and/or energy scales is found.
In the above example of coupled excitons and phonons,  the NN excitonic coupling energies $\beta$ typically exceed the vibrational energies $\nu,\omega$, which is due to the disparity of electronic and nuclear masses ~\citep{Lenz1951}.
In such cases, a promising way to overcome the \textit{curse of dimensionality} is to resort to hybrid quantum-classical molecular dynamics where only the light (fast) subsystem is treated quantum-mechanically while the classical approximation for the heavy (slow) subsystem is used.
Such approaches appear especially suitable for problems where a large local Hilbert space dimension $d$ is due to the latter subsystems being more complicated than those of Eq.~\eqref{eq:H_ph1}.
An example are conjugated polymer chains where the chromophoric sub-units are typically connected by a chain segment of several chemical bonds featuring a number of stretching, bending, and torsional degrees of freedom ~\citep{Binder2018,DiMaiolo2020}.

The simplest quantum-classical approach is given by mean field or Ehrenfest dynamics which rests on a separability \textit{ansatz}.
There, the state vector of the coupled system is assumed to be a single product of the two subsystem states which is also known as time-dependent Hartree method.
Moreover, the quantum (excitonic) states can be restricted to the Fock space of singly excited states $\sum_{i=1}^N a_i(t) b_i^\dagger |0\rangle$ with time-dependent, complex coefficients $a_i(t)$ and with $|0\rangle$ standing for the electronic ground state.
While this assumption neglects couplings to states bearing two or more excitons, it renders a TT-based approach for the excitons unnecessary.

For the example of the Hamiltonians of coupled excitons and phonons introduced in Sec.~\ref{sec:ham}, the evolution of the quantum sub-system (excitons) is governed by a Schr\"{o}dinger-type equation 
\begin{equation}
\mathrm{i}\frac{\mathrm{d}a_i}{\mathrm{d} t}=
\left[\alpha_i +\sigma_i (R_{i+1}-R_{i-1}) + W  \right] a_i 
+ \beta_{i-1} a_{i-1}+\beta_i a_{i+1}
\label{eq:schroedi}
\end{equation}
where $a_i(t)$ are the expansion coefficients of the excitonic state and where $W$ stands for the (classical) energy of the phonons. 
Further, the dynamics of the classical sub-system (phonons) is described in terms of a classical trajectory, $R(t)$, which is governed by a Newton-type equation
\begin{eqnarray}
m_i \frac{\mathrm{d}^2 R_i}{\mathrm{d}t^2}=
&=& -m_i \nu_i^2 R_i \nonumber \\
& & -\mu_{j-1} \omega_{j-1}^2 (R_i-R_{i-1}) - \sigma_{i-1} |a_{i-1}|^2 \nonumber \\
& & +\mu_j \omega_j^2 (R_{i+1}-R_i) + \sigma_{i+1} |a_{i+1}|^2
\label{eq:newton}
\end{eqnarray}
Note that here the two sub-systems given in Eqs~\eqref{eq:schroedi},~\eqref{eq:newton} are coupled to each other through terms proportional to the EPC constants $\sigma_i$ defined in the third row of Eq.~\eqref{eq:H_epc}.
For detailed discussions of the asymtotics and error estimates of the separabilty ansatz and/or the classical approximation see, e.g., Refs.~\citep{Bornemann1996,Davydov1985,Georgiev2019,Burghardt2021,Burghardt2022}.

The use of class \textit{QCMD} is shown in the following sample code lines
{\small\begin{verbatim}
   from wave_train.dynamics.qcmd import QCMD
   dynamics = QCMD(hamilton=hamilton, 
       num_steps=50, step_size=20, sub_steps=5,
       solver='sm', normalize=0) 
   dynamics.fundamental()
   dynamics.solve()
\end{verbatim}}
where \texttt{hamilton} has to be an object of class \textit{Coupled}, see Sec.~\ref{sec:ham:coupled}, or another class for bipartite systems provided by the user.
Note that in order to be used for Ehrenfest quantum-classical mechanics simulations, such classes have to provide  methods \textit{qu\_coupling} and \textit{cl\_coupling} returning the couplings of one sub-system to the respective other one, see also Fig.~\ref{fig:uml:ham}.
For numerically solving the QCMD scheme, there is a choice of numerical propagators implemented within the \textit{QCMD} class, such as a generalized Lie-Trotter (\texttt{'lt'}) and Strang-Marchuk method (option \texttt{'sm'} in the example code above), as well as the symplectic pickaback (\texttt{'pb'}) propagator~\citep{Nettesheim1996}. 

The choice of initial conditions for the quantum sub-system (e.g. excitons) is the same as in Sec.~\ref{sec:dyn:tdse} for class \textit{TDSE}, i.e., fundamental electronic excitations, with the possibility for Gaussian bell-shaped and sech-shaped superpositions thereof. 
Note that initial excitations of the classical sub-system (e.g. phonons) are at present not yet implemented.

\subsection{Class \textit{CEoM}: Classical Equations of Motion}
\label{sec:dyn:ceom}

Moreover, we have added a class for solving classical (Newton's or Hamilton's) equations of motion to the \Software package.
The motivation for this is to generate reference solutions for systems where a classical analogue to the quantum-mechanical Hamiltonian exists.
Hence, this class works, e.g., with objects of class \textit{Phonon}. 
According to the Ehrenfest theorem, quantum-mechanical expectation values of observables such as positions and momenta coincide with results from classical trajectories, as long as the vibrational Hamiltonian is a polynomial of order not higher than two, which is indeed the case for our harmonic model Hamiltonian~\eqref{eq:H_ph1}.
There, the positions are governed by the Newton-type equation~\eqref{eq:newton}, but without the $\sigma$ term for the EPC.  

The use of class \textit{CEoM} is illustrated in the following code lines
{\small\begin{verbatim}
   from wave_train.dynamics.ceom import CEoM
   dynamics = CEoM(hamilton=hamilton, 
       num_steps=50, step_size=20, sub_steps=5,
       solver='rk', normalize=0) 
   dynamics.coherent(displace=[1.0 if i == hamilton.n_site//2 else 0.0 for i in range(hamilton.n_site)])
   dynamics.solve() 
\end{verbatim}}
where \texttt{hamilton} is an object of class \textit{Phonon}, for a description see Sec.~\ref{sec:ham:phonon}, or another class provided by the user for a system for which the use of the classical approximation is justifiable.
Note that for use in classical mechanics simulations, such classes have to encompass additional methods for the calculations of forces and of classical potential and kinetic energy, see also Fig.~\ref{fig:uml:ham}.
For numerically solving the classical equations of motion, there is a choice of propagators implemented in class \textit{CEoM}, such as the Runge-Kutta (option \texttt{'rk'} in the example above) and the Velocity-Verlet (\texttt{'vv'}) scheme.
In addition, quasi-exact solutions for the harmonic vibrations are available which require additional Python methods to calculate the Hessian matrices of the potential and kinetic energy functions, see also Eq.~(25) of Ref.~\citep{Gelss2022b}.
In the sample code above, method \textit{coherent} of class  \textit{CEoM} is used to provide classical initial conditions equivalent to those of a coherent state of quantum harmonic oscillators with the displacement of the classical particles given by Eq.~\eqref{eq:displace}, here with $\langle R \rangle=1$ at the central site and $\langle R \rangle=0$ everywhere else.

\subsection{Class \textit{Load}: Loading data from a previous simulation}
\label{sec:dyn:load}

In addition to generating solutions of (stationary or dynamical) equations of motion as described in the subsections above, \Software also offers the possibility of loading previously generated solutions
If, for example, a TDSE simulation is run with option \texttt{load\_file=tdse\_1.pic}, essential data are stored in a Python pickle file by virtue of method \textit{save} in class \textit{Mechanics}, see Sec.~\ref{sec:dyn:super}.
Subsequently, this information is easily retrieved using class \textit{Load} 
{\small\begin{verbatim}
   from wave_train.io.load import Load
   dynamics = Load('tdse_1', 'pic') 
\end{verbatim}}
The created object contains not only expectation values of important physical observables
which can be used for automated analysis of series of runs, but also the TT representation of the last bound state (TISE) or the state at the last time step (TDSE) which allows for an easy restart of a simulation.
Finally, objects of class \textit{Load} contain also reduced density information which serve the purpose of creating a new (or different) animated visualization without having to perform another full simulation, see also the following section.

\section{Graphical output}
\label{sec:gra}
The ability to create rich graphical output is one of the hallmarks of simulations with the \Software software.
To meet the demand of users for rich and insightful graphical representations, the software package provides a set of default visualization classes.
They allow the user to track the progress and stability of computations at run time or to create graphical output of previously generated results by utilizing the \textit{Load} class, see Sec.~\ref{sec:dyn:load}.
After completion of a simulation, the plots are available not only as images (png file format) but also as animations (mp4 file format) which are created using the \textit{ffmpeg} tool.\cite{Tomar2006}

Classes for visualization are created based on a Dependency Injection (DI) scheme, with different visualization services 
being injected into the main class \textit{Visual}, that handles the execution order of the respective services.
Generally, visualizing the results of solving the equations of motion introduced in Sec.~\ref{sec:dyn} clusters into two independent services that can be separately added to the pipeline for creating visual output.
In the main service step the current quantum or classical state is visualized in a collection of subplots for each of the sites or in a single view along a discretized axis of site indices, in both cases shown in the left half of the generated figures.
Additionally, a second service can be added to monitor system properties, i.e., energy (TISE, TDSE, QCMD, CEoM), norm (TISE, TDSE, QCMD) and
autocorrelation function (ACF), $C(t)=\langle\psi(0)|\psi(t)\rangle$ (TDSE, QCMD), optionally displayed in the right half of the generated figures.

\begin{figure}
    \includegraphics[width=\linewidth, keepaspectratio]{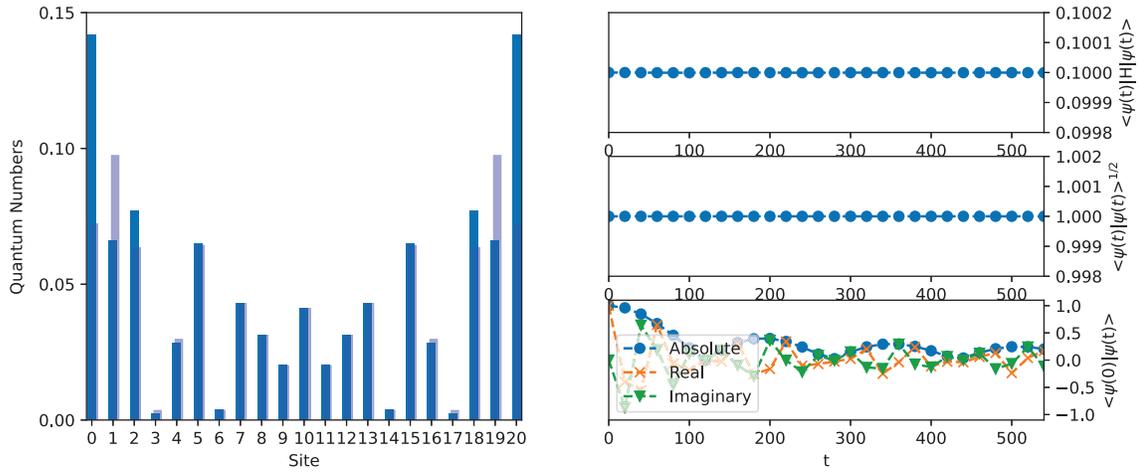}
    \caption{Quantum dynamics of excitons on a linear chain.
     Left panel: snapshot of quantum numbers for each of the sites.
     Right panel: evolution of the mean energy and the norm of the state vector versus time, as well as the autocorrelation function.}
    \label{fig:exci_tdse}
\end{figure}

The following code snippet illustrates the setup of an animation for visualizing the quantum dynamics of a single system, e.g., a chain of excitons as shown in Fig.~\ref{fig:exci_tdse}

{\small\begin{verbatim}

    from wave_train.graphics.factory import VisualTDSE

    graphics = VisualTDSE(
        dynamics=dynamics,
        plot_type='QuantNumbers',
        plot_expect=True,
        movie_file='tdse.mp4').create()
    graphics.solve()
\end{verbatim}}
Here, it is assumed that \textit{dynamics} is a previously created object of class TDSE, as described in Sec.~\ref{sec:dyn:tdse}.
This instance is then inserted into the factory constructor, which becomes responsible for internal logical checks, e.g., whether plot
type, \textit{hamilton} instance, and \textit{dynamics} instance are compatible.
% based on the system properties that are calculated at runtime.
The \textit{VisualTDSE} factory returns an instance of the \textit{Visual} class after a call to the \textit{create} method,
which will inject the respective services.
Visualizations of different \textit{dynamics} instances follows the same logic, with equivalent factory classes being provided for TISE, QCMD, and CEoM.
In the above code snippet, the service \textit{QuantNumbers} for displaying average quantum numbers for each of the sites has been selected and the toggle (\textit{plot\_expect}) for the visualization of system properties (expectation values of norm and energy, ACF) has been activated. 
The setup for high-level visualization of these observables is routed through the factory interface, that provides the factory classes for the different dynamics implemented in the \Software software (i.e. TISE, TDSE, QCMD, CEoM).
Please note that the new instance \textit{graphics} provides a proxy to start solving the Schrödinger equation, thus replacing the calls to \textit{dynamics.solve} in the code snippets given previously in Sec.~\ref{sec:dyn}.
Finally, specifying the \textit{movie\_file} keyword argument allows to create animated output in mp4 file format.

The visualization of the system state in the left half of the figures is based on the reduced density formalism.
Once calculated, the reduced density matrices for each site can be shown directly, or in the form of populations or averaged quantum numbers, positions and/or momenta.
For an overview of the different visualization options, see Tab.~\ref{tab:vis-services}.
That table also lists the special graphics services designed for use with bipartite systems, e.g. the coupled exciton-phonon systems described in Sec.~\ref{sec:ham:coupled}. 

Optionally, the system properties can be visualized in the right half of the figures.
These properties are directly calculated as overlaps or as expectation values by utilizing the tensor product as provided by \textit{scikit\_tt}.
Where possible, system properties are always separated into their individual contributions, e.g., for bipartite systems the state
space is visualized for the two sub-systems separately.
Furthermore, for CEoM simulations, the  system energy is split into  kinetic and potential energy contributions whereas for QCMD, energy contributions are decomposed into the contributions from the quantum and classical subsystem, as well as the energy pertaining to the quantum-classical coupling.

\begin{table}
    \begin{tabular}{llll}
    \hline \hline
        Service & System & Dynamics & Description \\
        \hline
        QuantNumbers    & Exciton, Phonon  & TISE, TDSE             & Mean quantum numbers \\ 
        \hline
        Populations     & Exciton, Phonon  & TISE, TDSE             & Populations of quantum states \\
        \hline
        DensityMat      & Exciton, Phonon  & TISE, TDSE             & Reduced density matrices \\
        \hline
        PhaseSpace      & Phonon           & TISE, TDSE, CEoM       & Mean trajectories in phase space \\
        \hline
        Positions2      & Coupled           & TISE, TDSE, QCMD       & \makecell[l]{Excitonic quantum number and\\lattice distortions as line plots} \\
        \hline
        QuantDisplace2  & Coupled           & TISE, TDSE, QCMD       & \makecell[l]{Excitonic quantum numbers and\\lattice distortions as bar plots} \\
        \hline
        QuantNumbers2   & Coupled           & TISE, TDSE,            & \makecell[l]{Excitonic and phononic \\quantum numbers as bar plots}  \\ \hline\hline
    \end{tabular}
\caption{Overview of the different visualization services and their cross-dependencies regarding dynamics and system. Upper four options for simple systems, lower three options for bipartite systems.}
\label{tab:vis-services}
\end{table}

\begin{figure}
    \includegraphics[width=\linewidth, keepaspectratio]{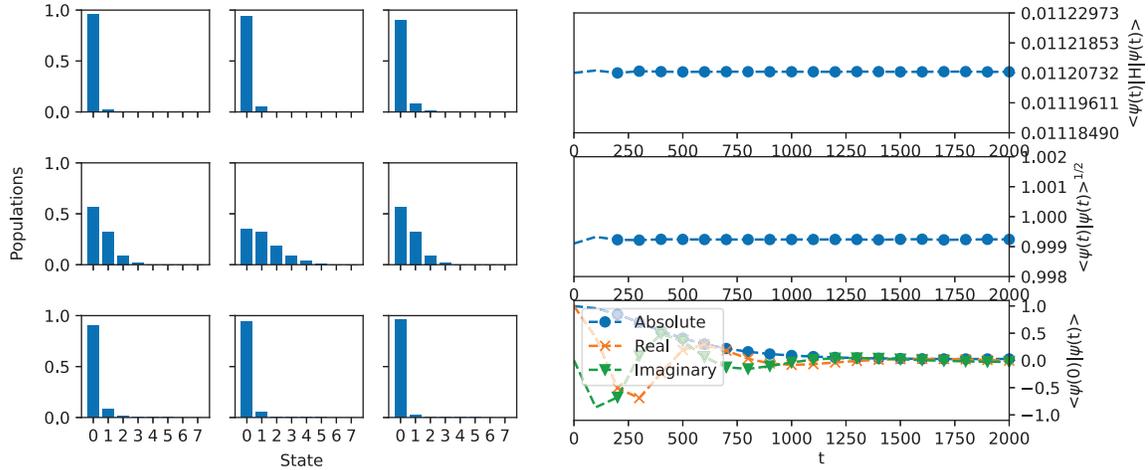}
    \caption{Quantum dynamics of phonons on a linear chain.
     Left panel: snapshots of populations of harmonic oscillator states for each of the sites, arranged in a row-wise manner.
     Right panel: Same as in Fig.~\ref{fig:exci_tdse}.}
    \label{fig:phon_tdse}
\end{figure}

\begin{figure}
    \includegraphics[width=\linewidth, keepaspectratio]{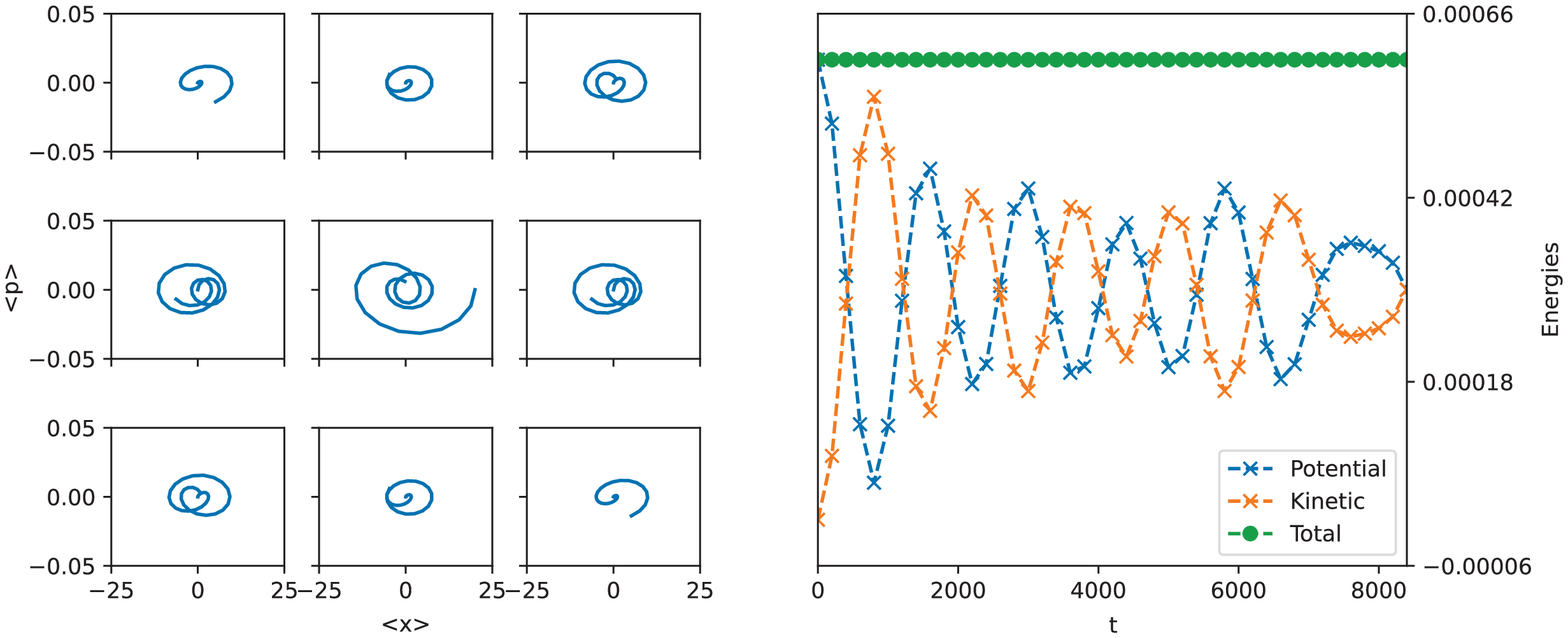}
    \caption{Classical dynamics of phonons on a linear chain.
     Left panel: trajectories in phase space for each of the sites. 
     Right panel: Total energy versus time, along with its decomposition in kinetic and potential contributions.}
    \label{fig:phon_ceom}
\end{figure}

\begin{figure}
    \includegraphics[width=\linewidth, keepaspectratio]{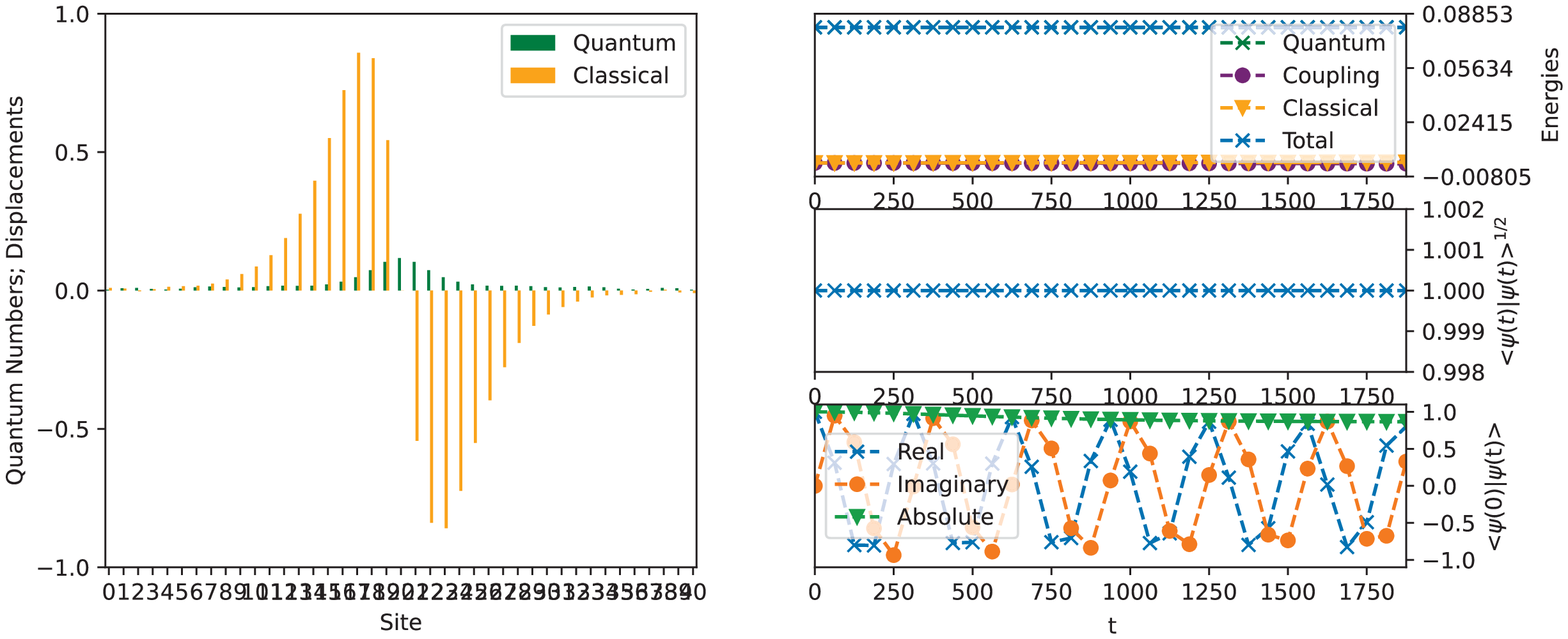}
    \caption{Quantum-classical dynamics of coupled excitons and phonons on a linear chain.
     Left panel: Snapshots of mean quantum numbers of the excitons (green bars) and vibrational displacements (scaled by 0.1) of the sites (orange bars). 
     Upper right panel: Total energy versus time, along with its decomposition in contributions of the quantum and the classical subsystem, as well as the quantum-classical coupling.
     Middle and lower right: Norm and autocorrelation function of the quantum subsystem only.}
    \label{fig:coup_qcmd}
\end{figure}

Typical graphical output from \textsc{WaveTrain} is illustrated and discussed for four selected cases.
\begin{itemize}
\item Fig.~\ref{fig:exci_tdse} is a visualization of the quantum dynamics of excitons on a linear chain of length $N=21$, with parameters from Sec.~\ref{sec:ham:exciton}.
The left half is showing a snapshot for $t=540$ after an initial excitation of the central site ($i=11$) only.
The semi-transparent bars show analytic solutions which are available for infinitely long chains of two-level systems with NN coupling only \citep{Kenkre1984}.
While analytic and numerical results agree well in the middle of the chain, there are considerable discrepancies near the edges of the chain, as expected.  
\item Fig.~\ref{fig:phon_tdse} shows the quantum dynamics of phonons on a linear chain of length $N=9$, with parameters from Sec.~\ref{sec:ham:phonon}.
The left panel represents a snapshots for $t=2000$ after an initial excitation of the central site ($i=4$) to a coherent state with $\langle R\rangle=50$.
For this value of the initial displacement, the representation of quantum state vectors in terms of 8 basis functions per site is almost large enough, with tiny deficiencies still visible in the deviation of the norm of the state vectors  from unity.
\item Fig.~\ref{fig:phon_ceom} visualizes the dynamics of phonons on a linear chain of length $N=9$, with parameters from Sec.~\ref{sec:ham:phonon}.
The left panel shows phase-space portraits for $0\leq t \leq 8400$ starting from an initial displacement with $\langle R\rangle=20$ of the central site ($i=4$) only. 
Note that for the quadratic Hamiltonian of Eq.~\eqref{eq:H_ph1}, resulting expectation values from quantum and classical dynamics coincide by virtue of the Ehrenfest theorem.
\item Fig.~\ref{fig:coup_qcmd} shows the quantum-classical dynamics of coupled excitons and phonons on a linear chain, with parameters from Sec.~\ref{sec:ham:coupled}.
The left part of the figure shows a snapshot at $t=1875$ after preparing an initial state with a sech-like distribution of an exciton peaked around the central site ($i=20$), see Ref.~\citep{Gelss2022a}, but without vibrational excitation.
Hence, this simulation shows the formation of a soliton or, more precisely, the onset of the dressing of an exciton with phonons in real time.
\end{itemize}

\section{Download and Installation}
\label{sec:install}
The \Software software is a pure Python3 package and can be readily installed from the PyPI package index
using pip. A command line installation of the \Software software can be achieved by issuing the following 
command in a terminal environment 
{\small\begin{verbatim}
    $ pip install wave_train
\end{verbatim}}
where pip installs into a Python3 installation with minimum version requirement 3.7.0.
The source code is publicly available in the Github repository \textit{PGelss/wave\_train} under 
the GNU General Public License v3.0.
For a developer installation of \Software, a specific version of \textit{sckit\_tt} may be required, 
which can be readily installed from the Github repository \textit{PGelss/scikit\_tt}. 
By default, \Software installs with the latest \textit{scikit\_tt} version.

\section{Conclusions and Prospect}
\label{sec:final}
In the present work, we have illustrated the use of \Software for rather simple models of excitons and phonons from our previous works \citep{Gelss2022a,Gelss2022b}.
However, it is straight-forward to apply our software to a variety of other quantum systems, as long as they are of a linear or cyclic chain-like topology with on-site and NN interactions only.
%  with or without periodic boundary conditions, and .
Obvious extensions of the models given above include exciton dynamics with more than two electronic states per site (e.g., singlet and triplet states) and/or anharmonic description of phonons.
Note that in the latter case one does not necessarily have to use the second quantization introduced in Eq.~\eqref{eq:H_ph2}.
It is also posssible to use, e.g., pseudo-spectral representations in coordinate space to discretize the vibrational degrees of freedom~\cite{Light2000,Schmidt2017}.
The flexible structure of the \Software package also allows for easy implementation of other types of quantum systems such as chains of spin systems (Ising or Heisenberg models), chains of molecular rotors~\citep{Serwatka2022}, or polarons in one-dimensional lattices~\citep{Devreese2009}.
In all those cases, one would have to design a new Python class for the underlying Hamiltonian which inherits from the super-class \textit{Chain}.
It is recommended for such a class to have a method \textit{\_\_init\_\_} dealing with the physical parameters of the Hamiltonian (class attributes) and a method \textit{\_\_str\_\_} generating a string for print output.
A mandatory ingredient of such a class is a method \textit{get\_SLIM} providing the $S,L,I,M$ matrices from which to construct the tensor cores~\citep{Gelss2017}, see also Eqs.~\eqref{eq:SLIM_1}, \eqref{eq:SLIM_2} in Sec.~\ref{sec:ham:slim}.

Moreover, the object-oriented architecture of the \Software package also supports a straight-forward addition of Python classes for further types of equations of motion.
An obvious choice is the Liouville-von Neumann equation (LvNE) adding dissipation and decoherence to quantum dynamics.
In that case, tensor trains will be used for the representations of the density matrices, and numerical solution of the LvNE will rest on the efficient ODE solvers available in the scikit\_tt toolbox, similar to our implementation of class \textit{TDSE} described in Sec.~\ref{sec:dyn:tdse}.
A frequently used alternative to the Ehrenfest or mean field quantum-classical approach implemented in class \textit{QCMD} is the surface hopping trajectory method, featuring stochastic hopping between different electronic states~\cite{Tully1990}.
In such a case, TT representations are not required, and a corresponding class should contain its own propagation methods, as is also the case for class \textit{QCMD} described in Sec.~\ref{sec:dyn:qcmd}.
Yet another option could be diffusive Langevin dynamics adding friction and stochastic forces to classical dynamics, thus extending the class \textit{CEoM}, see Sec.~\ref{sec:dyn:ceom}.
Note that the classes for these three examples will inherit from the respective super classes for fully quantum, mixed quantum-classical, and purely classic dynamics, as described in Sec.~\ref{sec:dyn:super}.
Moreover, when writing a new Python class for another type of dynamics, the following methods will have to be implemented:
In addition to a method \textit{\_\_init\_\_} for initialization and a method \textit{\_\_str\_\_} for print output, it is mandatory for such a class to encompass a method \textit{solve}.
That method calls \textit{start\_solve} used for initialization of the numerical solvers, e.g., propagation one step backwards in time which is required for the symmetric Euler scheme to solve the TDSE. 
Subsequently, for every time-step a method \textit{update\_solve} is called that actually carries out the propagation.
Finally, it should be mentioned that each of the dynamics classes needs to have (one or several) method(s) to generate an initial system state.

The \Software software package is hosted and further developed at the Github platform, along with the scikit\_tt toolbox for tensor train computations on which it is based.
Moreover, \Software is mirrored at the SourceForge platform, as a part of the \textsc{WavePacket} project for numerical quantum dynamics which is already in use for a number of years in several labs~\citep{Schmidt2017,Schmidt2018,Schmidt2019}.
That MATLAB software package also features quantum and mixed quantum-classical dynamics, but for general Hamiltonians, i.e., without the restriction to chain-like topologies.
The recently published version 7.0 of \textsc{WavePacket} contains a MATLAB class definition for the Hamiltonians of Eqs.~\eqref{eq:H_ex}, \eqref{eq:H_ph1}, \eqref{eq:H_epc} from the present work.
Hence, integration of \Software into the \textsc{WavePacket} project allows for a simple and direct comparability of results,
thus allowing to benefit from the easy usability and the advanced graphical capabilities of the latter one.
% thereby benefiting from the advanced graphical representations of wave functions in position, momentum, and partly also in Wigner distributions in phase space.
However, such comparisons will have to be limited to short chains up to $N\approx 3$ for TISE or $N\approx 6$ for TDSE because -  without the use of tensor train methods - \textsc{WavePacket} suffers from the \textit{curse of dimensionality}.

\begin{acknowledgments}
Funded by the Deutsche Forschungsgemeinschaft (DFG, German Research Foundation) under Germany's Excellence Strategy -- The Berlin Mathematics Research Center MATH+ (EXC-2046/1, project ID: 390685689) and by the CRC 1114 ``Scaling Cascades in Complex Systems'' funded by the Deutsche Forschungsgemeinschaft (project ID: 235221301).
Sebastian Matera (Fritz Haber Institute, Berlin) is acknowledged for insightful discussions.
\end{acknowledgments}

\section*{Author declarations}
\subsection*{Conflict of Interest}
No potential conflict of interest was reported by the authors.

\section*{Data Availability}
The Python scripts used to generate the results shown in Figs.~\ref{fig:exci_tdse}--\ref{fig:coup_qcmd} are openly available in the \textsc{Zenodo} repository at https://doi.org/10.5281/zenodo.7354077.

\newpage

\bibliography{Software_1} 

\end{document}